\documentclass[showpacs,twocolumn,pre,floatfix]{revtex4-1}
\usepackage{booktabs}
\usepackage{xcolor}
\usepackage{xtab}
\usepackage{amsmath}
\usepackage[utf8]{inputenc}
\usepackage{textcomp}
\usepackage{graphicx}
\usepackage{calc}

\newcommand{\Fex}{\ensuremath{F_{\rm ex}}}
\newcommand{\Vext}{\ensuremath{V_{\rm ext}}}
\newcommand{\dcf}[1][1]{\ensuremath{c^{(#1)}}}

\newcommand{\vect}[1]{\ensuremath{\boldsymbol{#1}}}
\newcommand{\av}[1]{{\ensuremath{\left\langle #1 \right\rangle}}}

\newcommand{\set}[1]{{\ensuremath{\{#1\}}}}

\newcommand{\vecr}{\ensuremath{\vect{r}}}

\newcommand{\pdiff}[3][]{\ensuremath{\frac{\partial^{#1}#2}{\partial#3^{#1}}}}

\newcommand{\rb}{{\vecr}}

\begin{document}

\author{J. Reinhardt}
\author{J.M. Brader}
\affiliation{Department of Physics, University of Fribourg, CH-1700 Fribourg, Switzerland}
\title{Dynamics of localized particles from density functional theory}

\begin{abstract}
A fundamental assumption of the dynamical density functional theory (DDFT) of colloidal systems is that 
a grand-canonical free energy functional may be employed to generate the thermodynamic driving forces. 
Using one-dimensional hard-rods as a model system we analyze the validity of this key assumption 
and show that unphysical self-interactions of the tagged particle density fields, arising from coupling 
to a particle reservoir, are responsible for the excessively fast relaxation predicted by the theory. 
Moreover, our findings suggest that even employing a canonical functional would not lead to an 
improvement for many-particle systems, if only the total density is considered.  We present several 
possible schemes to suppress these effects by incorporating tagged densities.
When applied to confined systems we demonstrate, using a simple example, that DDFT neccessarily leads 
to delocalized tagged particle density distributions, which do not respect the fundamental geometrical 
contraints apparent in Brownian dynamics simulation data. 
The implication of these results for possible applications of DDFT to treat the glass transition are 
discussed. 
\end{abstract}


\maketitle

\section{Introduction}
During the past decade of liquid-state research the classical dynamical density functional theory (DDFT) has 
proven to be a versatile and reliable tool for describing the dynamics of interacting colloidal particles in a 
wide variety of situations. 
Building upon the success of equilibrium DFT \cite{evans79,evans92}, the dynamical theory enables first-principles 
calculation of the inhomogeneous density $\rho(\rb,t)$ generated in response to a time-dependent external 
potential field $V_{\rm ext}({\bf r},t)$ \cite{marconitarazona99}. 
Within the DDFT framework a nonvanishing particle flux arises solely from gradients in a local chemical 
potential $\mu({\bf r},t)$, derived from an equilibrium free energy functional.  
The theory has been successfully applied to many problems, including relaxation to equilibrium from a given 
nonequilibrium initial state \cite{marconitarazona99}, the early and intermediate stages of spinodal decomposition \cite{archer04} and 
systems for which the time-dependence of $V_{\rm ext}({\bf r},t)$ drives the system into nonequilibrium steady or 
stationary states \cite{penna_tarazona,penna2,tarazona_marconi08,rex09,archer11}. 
In recent years the original formulation of DDFT has been extended to treat both systems and situations 
of ever increasing complexity. 
These more recent developments have lead to a better understanding of the influence of both nonpotential fields 
(e.g. mechanical \cite{kr1,kr2,bk,bk2}, thermal \cite{melchionna07,tarazona_marconi08}) and particle 
geometry (see e.g. \cite{rex_wenksink07}) on diffusive colloidal dynamics. 

Despite the undeniable success of the DDFT in robustly capturing the qualitative features of $\rho(\rb,t)$ 
for many problems of interest, the theory rests upon two fundamental assumptions, both of which 
remain to be either systematically evaluated or improved upon. 
The first of these is the so-called adiabatic approximation, namely the assumption that 
the two-body correlations may be calculated from the instantaneous one-body density using 
equilibrium statistical mechanical relations. 
The second major assumption is that the nonequilibrium chemical potential 
$\mu({\bf r},t)$ generating the particle dynamics can be 
identified with the functional derivative of a grand-canonical free energy. 
Combining these two approximations yields a closed equation for the one-body density, which does 
not require explicit knowledge of the higher-order correlations. 

In the present work we investigate the validity of employing a grand-canonical density functional to 
treat many-body effects in DDFT. 
Problems can be anticipated in confined systems with small particle number, for which the 
choice of ensemble strongly influences the equilibrium density profiles. 
More generally, use of the grand-canonical ensemble becomes questionable 
for situations where the density field is strongly localized in space and contain only few particles. This can occur either as a direct consequence of 
minima in $V_{\rm ext}({\bf r},t)$ or as a transient which may occur along the natural diffusive 
trajectory of $\rho(\rb,t)$ through the space of density 
functions. 
Transient localization occurs quite naturally when considering the time-evolution of the density 
from sharply defined initial conditions, for which the positions of all $N$-particles, or a subset thereof,  
are known precisely. 
The individual density peaks of particles sharply located at $t=0$ remain well separated for short times and 
are normalized to unity (each peak contains a single particle). 
%However, taking a complete equilibrium statistical average at each timestep of the relaxation, as is done 
%in a DDFT calculation, neglects the kinetic constraints on the particle positions and introduces 
%ensemble dependent relaxation dynamics.  

The above considerations become even more pertinent when considering potential applications of grand-canonical 
DDFT to describe dynamic arrest and glass formation. 
By tagging the density field of a single particle in a dense hard-sphere liquid it has recently been shown 
\cite{hopkins2010} that the tagged density (the self part of the van Hove function) can exhibit a two step 
relaxation within DDFT, leading to dynamic arrest at sufficiently high volume fractions. 
Similar behaviour was also found for particles interacting via a Gaussian pair potential \cite{hopkins2007}.
However, due to the use of an approximate free energy functional it remains unclear whether the observed 
slow dynamics 
arises purely from the presence of (possibly spurious) metastable minima in the free energy, or whether it is  
a genuine physical prediction of the DDFT which would persist even when using an exact equilibrium functional.
The work of Hopkins {\em et al.} \cite{hopkins2007,hopkins2010} raises a fundamental question: 
Can a theory which involves only the one-body density field capture, even in principle, the transition to 
a nonergodic state?

A major difficulty in answering the above question for realistic glass forming systems in two or three dimensions is the absence of an exact equilibrium free energy functional. 
It thus becomes difficult to disentangle errors in the dynamical theory from those induced by an approximate 
free energy.  
For this reason we focus on the simplest nontrivial model for which the grand-canonical free energy functional 
is known exactly, namely a system of hard-rods in one dimension 
\cite{percus76,percus83}. 
Despite its simplicity, the hard-rod model has one feature in common with glassy states occuring in 
higher dimensional systems: 
Both exhibit a partitioned phase space in which physically meaningful averages must be taken over a subset of phase space
that is dynamically accessible. 
In glasses this partitioning occurs spontaneously as a thermodynamic control parameter crosses 
some critical value, whereas for hard-rods the reduced phase space is always present, simply as a result of 
ordering the particles on a line. 
The confinement of a given rod by its neighbours to the left and right presents the prototypical glassy 
`cage' and serves as a useful reality check for theories aiming to describe dynamical arrest in 
higher dimensions.  

The paper will be structured as follows: 
In Section \ref{sec:fundamentals} we first will outline the microscopic dynamics under consideration,  
review the standard formulation of DDFT and define the functional to be employed. 
In Section \ref{singlerod} we consider the diffusion of a single rod and various methods by which 
grand-canonical contributions can be eliminated. 
In Section \ref{manyrods} we identify the importance of tagging the individual density profiles. 
In Section \ref{numerical_results} we present numerical results for various test cases invlolving several 
interacting rods. 
Finally, in Section \ref{discussion} we provide a discussion of our results and and identify some 
future challenges.

\section{Fundamentals}\label{sec:fundamentals}

\subsection{Microscopic dynamics}
We consider a system of $N$ colloidal particles in a time-dependent external potential 
$V_{\rm ext}({\bf r},t)$ interacting via the spherically symmetric pair potential 
$\phi(r)$. 
The total potential energy is given by 
\begin{eqnarray}
U_N( \{ \rb_i \},t)
=\sum_{i}V_{\rm ext}(\rb_i,t) + \sum_{i<j}\phi(|\rb_i-\rb_j|).
\label{potential}
\end{eqnarray} 
As the individual colloidal trajectories are stochastic it is appropriate to adopt a
probabilistic description of the particle motion. The probability distribution of particle positions is 
described by the Smoluchowski equation \cite{dhont}
\begin{eqnarray}
\hspace*{1.1cm}\frac{\partial P(t)}{\partial t} + 
\sum_{i} \nabla_i\cdot {\bf j}_i =0,
\label{smol}
\end{eqnarray} 
where $P(t)\equiv P(\{ \rb_i\} ,t)$. 
The fact that Eq.(\ref{smol}) takes the form of a continuity equation expresses the conservation 
of particle number.
Neglecting the influence of solvent induced hydrodynamic interactions (see \cite{brader10} for a discussion of 
the implications of this approximation) the probability flux of particle $i$ is given by
\begin{eqnarray}
{\bf j}_i=
- D_{0}(\nabla_i - \beta\,{\bf F}_i) P(t),
\label{individualflux}
\end{eqnarray}
with $\beta=1/k_BT$. 
The force ${\bf F}_i$ on particle $i$ is generated from the total potential energy (\ref{potential})
according to ${\bf F}_i=-\nabla_i U_N$.

\subsection{Dynamic Density Functional Theory}\label{ssec:ddft}
Integration of Eq.(\ref{smol}) over all but one of the particle coordinates leads to an exact coarse-grained 
expression for the one-body density profile
\begin{eqnarray}\label{eq:eom_rho1}
\frac{\partial \rho({\bf r}_1,t)}{\partial t} = -\nabla_1\!\cdot {\bf j}(\rb_1,t), 
\end{eqnarray}
where the particle flux involves an integral over the nonequilibrium two-body density
\begin{eqnarray}\label{flux}
{\bf j}(\rb_1,t)&=&
\Gamma k_B T\,\nabla_1 \rho({\bf r}_1,t)  
\;+\;\Gamma\rho({\bf r}_1,t) \,\nabla_1 \Vext({\bf r}_1,t) \notag\\
&&+\;\Gamma\int \!\mathrm{d}\vecr_2 \nabla_1 \phi(r_{12})\,\rho^{(2)}({\bf r}_1,{\bf r}_2,t),
\end{eqnarray}
with mobility $\Gamma=D_0/k_BT$.
Eq.(\ref{eq:eom_rho1}) is the first in a hierarchy of equations for the $n$-body density.  
In equilibrium the flux is identically zero and Eq.(\ref{eq:eom_rho1}) reduces 
to the first member of the Yvon-Born-Green hierarchy \cite{hansenmcdonald86}. 

%In order to arrive at a closed expression for $\rho({\bf r},t)$ it is neccessary to approximate the integral 
%term in Eq.(\ref{flux}). 
%A simple first approach could be to employ a superposition approximation for the two-body density 
%$\rho^{(2)}({\bf r}_1,{\bf r}_2,t)\approx \rho({\bf r}_1,t)\rho({\bf r}_2,t)$.  
%However, the experience from equilibrium studies is that superposition is accurate only at low densities 
%(c.f. the Vlasov equation \cite{hansenmcdonald86}) and gives a poor account of the pair correlations, a 
%state of affairs which seems unlikely to improve when turning to address nonequilibrium situations. 

The integral term in (\ref{flux}) may be approximated as an explicit functional of the one-body density 
using the methods of equilibrium DFT. 
The Helmholtz free energy is split into three contributions 
\begin{eqnarray}\label{splitup}
\mathcal{F}=\mathcal{F}_{\rm id} + \mathcal{F}_{\rm ex} 
+ \int d\rb_1\;V_{\rm ext}(\rb_1)\,\rho(\rb_1), 
\end{eqnarray}
where $\Lambda$ is the thermal de Broglie wavelength and the ideal gas free energy is known exactly
\begin{eqnarray}\label{ideal}
\mathcal{F}_{\rm id}[\,\rho(\rb_1)]=\int d{\bf r}_1\, \rho(\rb_1)[\,\ln(\Lambda^3\rho(\rb_1))-1\,].
\end{eqnarray}
The excess free energy functional $\mathcal{F}_{\rm ex}[\,\rho({\bf r})]$ contains all information regarding the interparticle interactions and is connected to the average interaction force via the grand-canonical 
sum rule \cite{evans79}
\begin{eqnarray}\label{sumrule}
\int\!d\rb_2\, \nabla_1\phi(r_{12}) \rho(\rb_2|\rb_1)
= \nabla_1\frac{\delta \mathcal{F}_{\rm ex}[\rho({\bf r}_1)]}{\delta \rho(\rb_1)},
\end{eqnarray}
where we have introduced the conditional distribution $\rho(\rb_2|\rb_1)\equiv\rho^{(2)}(\rb_1,\rb_2)/\rho(\rb_1)$, 
i.e. the average number density at $\rb_2$ given a particle is fixed at $\rb_1$.

The assumption that Eq.(\ref{sumrule}) remains valid in nonequilibrium is the so-called adiabatic 
approximation. It is equivalent to assuming that the pair 
density $\rho^{(2)}(\rb_1,\rb_2,t)$ relaxes instantaneously 
to the equilibrium pair-density corresponding to the current one-body density $\rho(\rb_1,t)$.
The total particle flux (\ref{flux}) may thus be written in the form
\begin{eqnarray}\label{gradchem}
{\bf j}(\rb_1,t) = -\Gamma\rho(\rb_1,t)\nabla_1\mu(\rb_1,t), 
\end{eqnarray}
where the nonequilibrium chemical potential is given by 
\begin{eqnarray}\label{chem}
\mu(\rb_1,t)=\frac{\delta \mathcal{F}[\,\rho(\rb_1,t)]}{\delta \rho(\rb_1,t)}.
\end{eqnarray}
Combining Eqs.(\ref{eq:eom_rho1}), (\ref{gradchem}) and (\ref{chem}) yields the familiar form of the DDFT 
equation of motion
\begin{eqnarray}
\frac{\partial \rho(\rb_1,t)}{\partial t} = \nabla_1\cdot\Bigg[ \Gamma \rho(\rb_1,t) \nabla_1\frac{\delta \mathcal{F}[\rho(\rb_1,t)]}{\delta \rho(\rb_1,t)} \Bigg].
\label{stDDFT}
\end{eqnarray}
In standard applications of DDFT the free energy $\mathcal{F}$ entering (\ref{stDDFT}) is a grand-canonical 
quantity. 
Eq.(\ref{stDDFT}) thus describes the evolution of the density between two grand-canonical states, subject to the 
constraint that the {\em average} particle number is conserved.

For the present work the DDFT equation is solved numerically using finite difference techniques. 
The time integration is performed using the Euler method and the spatial dervatives with central differences. 
The integrations required for the convolutions are evaluated using the trapezoidal rule, which in one dimension 
can be calculated very efficiently due to the finite range of the weight functions.

\subsection{The Percus Hard-rod functional}
For our numerical investigations we will employ the exact excess Helmholtz free energy functional 
of hard-rods in one dimension \cite{percus76,percus83}. 
For an $m$ component mixture of rods the functional is given by
\begin{eqnarray}\label{percus_functional}
\mathcal{F}_{\rm ex}[\,\rho(x)] =\int_{-\infty}^{\infty} \!\!dx\;\; \Phi(\{n_{\alpha}\}),
\end{eqnarray}
where the free energy density, $\Phi=-n_0(x)\ln(1-n_1(x))$, is a function of a set of weighted densities 
\begin{equation}
n_{\alpha}(x)=\sum_{i=1}^{m}\int_{-\infty}^{\infty} \!\!dx'\; \rho_i(x')
\,\omega_i^{(\alpha)}(x-x'), 
\end{equation}
where $\rho_i(x)$ is the density profile of species $i$.
The weight functions are characteristic of the particle geometry, with  
$\omega^{(1)}_i(x)\!=\!\Theta(|\,x|-d_i/2)$ and  
$\omega^{(0)}_i(x)\!=\!\delta(|\,x|-d_i/2)/2$,
where $d_i$ is the length of rod species $i$.

\section{Single particle diffusion}\label{singlerod}

\begin{figure}
\hspace*{-0.5cm}
\includegraphics[width=8cm]{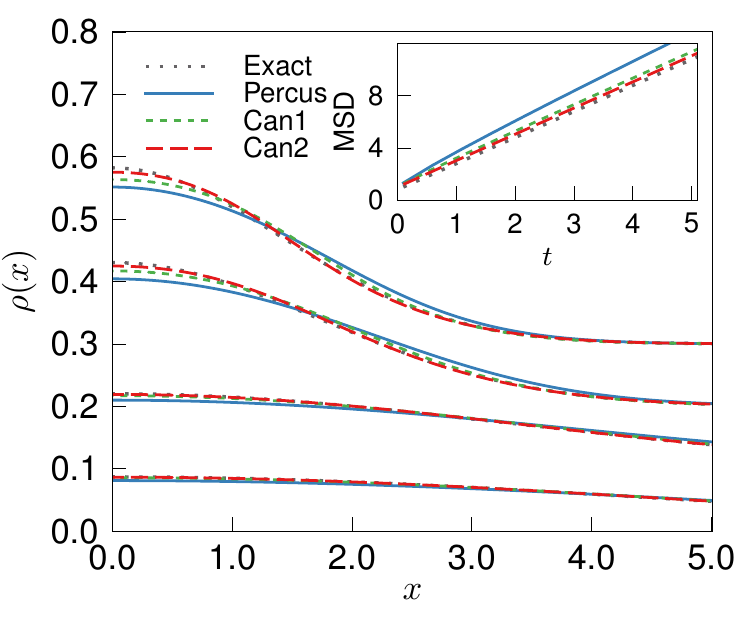}
\caption{\label{fig:one-particle-diff} (Color online)
The time evolution of density of a single particle for times $t=0.5,1.0,5.0,10.0$ (in units of $d^2/D_0$). 
The length scale is set by the rod length $d$.
We compare the exact result (grey dotted) with standard DDFT using the exact Percus functional (blue solid) and both first order (green short dashed) and second order (red long dashed) canonical 
corrections. The inset shows the corresponding mean squared displacement. }
\end{figure}

We begin by considering a single rod, whose diffusion is governed by the
diffusion equation, resulting in a Gaussian density distribution. The diffusion
equation can be recovered from the DDFT equation of motion \eqref{stDDFT} in
the case of vanishing interactions, i.e.  $\mathcal{F}_{\rm ex} = 0$.
In Fig.\ref{fig:one-particle-diff} we compare the results of DDFT 
(\ref{stDDFT}) using the Percus functional with the exact Gaussian result for the relaxation of the density profile 
from the sharp initial condition $\rho(x)=\delta(x)$. 
The inset to Fig.\ref{fig:one-particle-diff} shows the 
corresponding mean squared displacement (MSD) as a function of time. 
The DDFT reproduces neither the expected Gaussian form of the density profile nor the linear 
increase of the MSD as a function of time. 
Only at long times does the slope of the MSD approach unity as the local density becomes very small for all $x$ and 
the ideal gas term starts to dominate the free energy (\ref{splitup}).  
The deficiency of the theory lies in the nonvanishing contribution from the excess free energy, which 
leads to an effective self interaction of the density field and consequent enhanced relaxation rate.  
As originally pointed out by Marconi and Tarazona \cite{marconitarazona99}, employing a grand 
canonical functional effectively puts the system into contact with an unphysical particle reservoir, 
such that even for $\langle N\rangle=1$ the density distribution contains additional contributions 
from states with $N=0,2,3,\cdots$. The interaction term thus does not vanish, as it should for a single 
particle, because states with $N>1$ naturally incorporate interparticle interactions. 
The first step towards an improved theory is thus to eliminate, or at least reduce, the interaction between 
the physical particle and the reservoir.

\subsection{Canonical correction}\label{can_correction}
In Refs.\cite{Gonz'alez1997} and \cite{Gonz'alez1998} Gonz\'alez {\em et al.} proposed a scheme by which 
canonical equilibrium density profiles can be expanded in terms of grand-canonical averages. The method 
can be interpreted as a formal expansion of the canonical density profile in inverse powers of 
$\langle N\rangle$. 
Using this expansion, it was found possible to systematically correct the grand-canonical density profiles 
predicted by equilibrium DFT to achieve an improved description of a hard-sphere fluid confined in a small 
spherical cavity. 
Generalizing the arguments presented in \cite{Gonz'alez1997,Gonz'alez1998}, the canonical average of an 
arbitrary function of the particle positions $A(\{\rb_i\})$ may be expressed in terms of grand-canonical 
averages. 
To second order the expansion is given by 
\begin{eqnarray}\label{eq:avtrans}
\av{A}^{\rm c} = \av{A} + f_1(A) + f_2(A) + \cdots, 
\end{eqnarray}
where $\langle\cdot\rangle$ is a grand-canonical average and $\langle\cdot\rangle^c$ is a canonical average.    
The correction terms are given by
\begin{eqnarray}
f_1(A) =& - \frac{1}{2} \av{(N - \av{N})^2} \pdiff[2]{\av{A}}{\av{N}}\notag\\
f_2(A) =& - \frac{1}{6} \av{(N - \av{N})^3} \pdiff[3]{\av{A}}{\av{N}}\notag\\
		&-\frac{1}{2} \av{(N - \av{N})^2} \pdiff[2]{f_1}{\av{N}},\notag 
\end{eqnarray}
Due to the scaling of the partial derivatives in (\ref{eq:avtrans}) the terms $f_1$ and $f_2$ are of order 
$\av{N}^{-1}$ and $\av{N}^{-2}$, respectively. 
The utility of the expansion (\ref{eq:avtrans}) lies in its rapid convergence, even for very small 
values of $\langle N\rangle$ \cite{Gonz'alez1997,Gonz'alez1998}. 

Returning to the diffusion of a single particle, we now seek to use (\ref{eq:avtrans}) to suppress unwanted grand-canonical contributions to the dynamics of $\rho(\rb,t)$. 
At any given time we can use the instantaneous density profile 
$\rho(\rb,t)$ to construct an effective external potential 
\begin{align}
V_{\rm eff}(\rb,t) = -\beta^{-1}\ln(\rho(\rb,t)) - \dcf[1][\rho(\rb,t)] + \ln(z),
\label{eq:Veff}
\end{align}
where the fugacity term $\ln(z)$ is a physically irrelevant constant. 
The one-body direct correlation function 
\begin{eqnarray}
c^{(1)}(\rb_1)=-\frac{\delta \beta\mathcal{F}_{\rm ex}[\rho\,]}{\delta\rho(\rb_1)}
\end{eqnarray}
is evaluated using the instantaneous density.  
When employed in an {\em equilibrium} calculation, the potential $V_{\rm eff}(\rb,t)$ will, by construction, 
yield the equilibrium results for $\rho(\rb,t)$ and all other quantities accessible to DFT. 
The adiabatic approximation is equivalent to assuming that the higher-order nonequilibrium 
correlations are equal to the higher-order equilibrium correlations calculated in the presence of 
$V_{\rm eff}(\rb,t)$. 

We now define $I(\rb_1,t)$ as the average interaction force in the grand-canonical ensemble 
\begin{eqnarray}
I(\rb_1,t)\equiv 
\int\!d\rb_2\, \nabla_1\phi(r_{12}) \rho^{(2)}_{\rm gc}(\rb_1,\rb_2)
\end{eqnarray}
where the subscript (gc) makes explicit the fact that the 
pair-density inside the integral is a grand-canonical average. 
Using (\ref{sumrule}) yields
\begin{eqnarray}\label{av_force}
I(\rb_1,t)=-k_BT\rho(\rb_1,t)\nabla_1 c^{(1)}(\rb_1,t) 
\end{eqnarray}
Expressed in this way, it is rather natural to employ the expansion (\ref{eq:avtrans}) 
to approximate the required integral in terms of known grand-canonical quantities 
\begin{eqnarray}\label{canonical_approx}
\int\!d\rb_2\, \nabla_1\phi(r_{12}) \rho^{(2)}(\rb_1,\rb_2)&=& 
I(\rb_1,t) + f_1(I(\rb_1,t)) 
\notag\\ 
&&+\, f_2(I(\rb_1,t)) + \cdots
\end{eqnarray}
When calculated to all orders, the corrected interaction force should be zero for a single particle. 
We now proceed to explicitly calculate the first two correction terms in (\ref{canonical_approx}). 
The practical implementation of our scheme is as follows: At each time step in the 
numerical integration of (\ref{smol}) the effective potential (\ref{eq:Veff}) is constructed from the instantaneous density. The partial derivatives 
required to evaluate the functions $f_n$ appearing in (\ref{canonical_approx}) are then calculated 
numerically by performing equilibrium DFT calculations in the presence of a fixed $V_{\rm eff}(\rb_1,t)$. 
The series (\ref{canonical_approx}) is then evaluated to the desired order and used to generate the density 
distribution at the next timestep. 
Grand-canonical contibutions to the time-evolution of the density $\rho(\rb,t)$ can thus be 
suppressed on-the-fly in order to provide a more realistic description of the trajectory through the space 
of density functions.

In Fig.\ref{fig:one-particle-diff} we show the time-evolution of the density of a single particle 
corrected using (\ref{canonical_approx}) to both first and second order. 
The series converges very rapidly and the second order results are virtually indistinguishable from 
the exact Gaussian function, despite the fact that $\langle N\rangle=1$. 
Similar rapid convergence has been observed in the 
static case \cite{Gonz'alez1998}. 
%We note that a calculation similar to that presented here could also have been carried out using 
%the canonical Helmholtz free energy functional proposed in \cite{white2000} (and which agrees to first 
%order with the expansion (\ref{canonical_approx}) \cite{white2000b}). However, the numerical effort 
%involved in implementing the canonical functional is considerable and would offer no greater insight 
%than using Eq.(\ref{canonical_approx}).  

\section{Multiple particle diffusion}\label{manyrods}
We have now established that a systematic suppression of the grand-canonical contributions 
to the dynamics indeed leads to improved results, at least for the trivial case of a single particle. 
However, application of the same procedure to systems with $N>1$ reveals an additional 
complication.

\begin{figure}
\begin{center}
\includegraphics{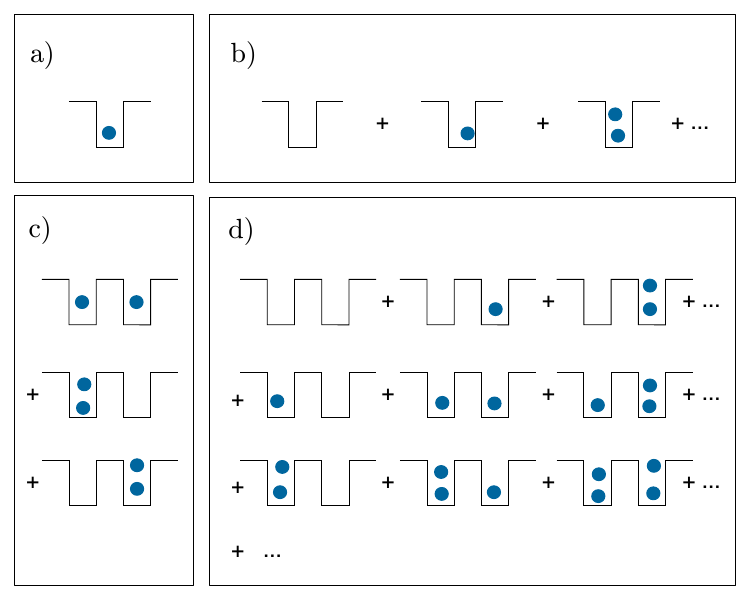}
\end{center}
\caption{\label{fig:ensembles} (Color online) 
Schematic illustration of the microstates that
contribute to canonical and grand-canonical averages for particles in one and
two infinite potential wells. (a) Canonical average $N=1$ (b) Grand-canonical
average $\av{N} = 1$ (c) Canonical average $N=2$ (d) Grand-canonical average
$\av{N}=2$. The second and third microstate in (c) are unphysical in the context
of DDFT for localized particles and lead to a overestimation of the
interactions.}

\end{figure}

Consider first the equilibrium average for a single infinite potential well in the canonical ensemble with 
$N = 1$ and in the grand-canonical ensemble with $\av{N} = 1$ (see figure
\ref{fig:ensembles}a and b). The average value of a quantity is the average over the
value of this quantity for all microstates appropriate to the given ensemble. 
As a result, the average density
profile for canonical and grand-canonical case are different because of the
additional microstates arising from coupling to an external particle reservoir.

The situation is similar to DDFT applied to transiently localized particles which 
have had not had enough time to diffuse far away from their starting position. 
The average interaction force is a grand canonical average, and the potential
wells discussed above are manifest in the effective external potential (\ref{eq:Veff}). 
In the grand-canonical case the microstates with more than one particle in the
well give rise to a nonvanishing interaction, which leads to the erroneous MSD.
The canonical corrections suppress these unwanted microstates and the resulting
treatment is closer to the canonical case.

\begin{figure}
\hspace*{-0.5cm}
\includegraphics[width=8cm]{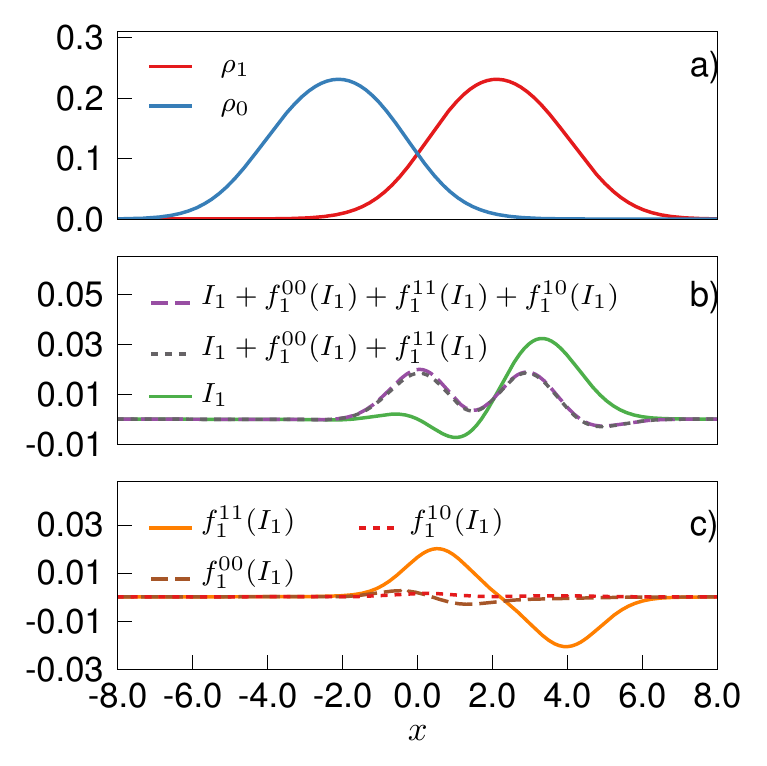}
\caption{\label{mix_corrections} (Color online) 
(a) Test case of two overlapping Gaussian density peaks. 
(b) The grand-canonical average force $I_1=-k_BT\rho_1(\rb_1)\nabla_1 c_1(\rb_1)$ (green solid line) acting on the density field
$\rho_1(\rb_1)$ (right peak). Negative values indicate a force pointing to the left.  
Also shown are the full first order canonical correction to the average force 
(long dashed line) and the same quantity without the cross correction $f_1^{01}$ (short dashed line).  
(c) The three terms contributing to the first order canonical correction.  
The length scale is given by the rod length.
}
\end{figure}

Next consider two infinite potential wells. The canonical correction series can
again suppress the microstates with more than two particles in total. But the
canonical ensemble still includes microstates with two particles in the same
well. 
In the context of DDFT for two localized particles these microstates again lead to an unphysical interaction force, which does not vanish even if the two density peaks are far away from each other. The interaction force on a particle resulting from number fluctuations in the $V_{\rm eff}(\rb,t)$ well generated by the particle can therefore be interpreted as a self interaction.

For larger numbers of particles and wells each well is coupled to a reservoir formed by the other wells. 
As $N$ increases the canonical and grand-canonical average become increasingly similar and the 
canonical correction decreases in magnitude, rendering it useless for the suppression of the self 
interaction.

What is required is a method to prevent the averaging procedure from including microstates states with more than one 
particle per localized density peak.

\subsection{Tagged particle approach}
An improved description can be achieved by viewing the $N$-particle system as an $N$-component 
mixture, in which each component corresponds to a single particle. This allows to precisely locate particles, even after their densities started overlapping.
%The individual density fields provide more information, but the price paid in formulating the system as a mixture is that the computational resources 
%required for implementation scale with particle number, thus limiting practical applications 
%to small systems. 
%For large numbers of (nonlocalized) particles the Standard DDFT should 
%perform well, and so we would not propose application of our method to large systems.

The DDFT for an arbitrary $m$-component mixture is a set of $m$ coupled equations 
\begin{eqnarray}
\frac{\partial \rho_i(\rb_1,t)}{\partial t} = \nabla_1\cdot\Bigg[ \Gamma \rho_i(\rb_1,t) \nabla_1\frac{\delta \mathcal{F}[\{ \rho_i\}]}{\delta \rho_i(\rb_1,t)} \Bigg],
\label{stDDFTmix}
\end{eqnarray} 
where $i$ labels the species. 
By tagging each individual particle Eq.(\ref{stDDFTmix}) constitutes a set of
$m$ partial differential equations in $D+1$ dimensions ($D$ space dimensions
and time) for the time-dependent one-body density fields.  On the other hand,
the Smoluchowski equation \eqref{smol} for such a system is a partial
differential equation in $D^m + 1$ dimensions for the probability distribution.
A numerical solution of the Smoluchowski eqution is therefore very demanding
and only possible for very few particles.

Secondly, in real applications one is not forced to tag each particle individually and (\ref{stDDFTmix}) offers 
complete flexibility as to which subsets of particles are associated with distinct species. For example, in the dynamic 
test particle calculations of \cite{hopkins2010} only a single particle was tagged and the rest of the system treated as 
a second component. 

%For the two particles under consideration we consider using Eq.(\ref{stDDFTmix}) to generate the time evolution 
%of the density profiles, but using a two component canonical correction procedure at each time step to reduce 
%grand-canonical contributions to the interaction $-k_BT\rho_i(\rb_1,t)\nabla_1 c_i(\rb_1,t)$. 

The application of the full correction series with a infinite number of terms to such a tagged system corresponds to a physically sensible average, that does neither suffer from particle fluctuations, nor from the combinatorial effects discussed above.
However, for large numbers of species and higher orders the canonical correction series becomes rather unwieldy,  
but progress can be made by only keeping some of the terms. This can be demonstrated by considering a pair of interacting particles, each 
carrying a distinct species label. 

To first order, the canonical correction for such a binary mixture is given by 
\begin{eqnarray}\label{can_correct_mix}
\av{A}^{\rm c} = \av{A}^{\rm gc} + f_{1}^{11}(A) + f_{1}^{00}(A)+ f_{1}^{01}(A) , 
\end{eqnarray}
where the correction terms are given by
\begin{eqnarray}\label{coefficients}
f_{1}^{11}(A) =& - \frac{1}{2} \frac{\partial \av{N_1}}{\partial \beta\mu_1} \pdiff[2]{\av{A}}{\av{N_1}}\notag\\
f_{1}^{00}(A) =& - \frac{1}{2} \frac{\partial \av{N_0}}{\partial \beta\mu_0} \pdiff[2]{\av{A}}{\av{N_0}}\notag\\
f_{1}^{01}(A) =& - \frac{\partial \av{N_0}}{\partial \beta\mu_1} \frac{\partial^2 \av{A}}{\partial \av{N_0} \partial \av{N_1}}.\notag
\end{eqnarray}
Higher order terms are of similar form, but involve higher derivatives.

In analogy with the single particle calculation presented in Section \ref{singlerod}, the first order 
correction (\ref{can_correct_mix}) could be used to determine the time evolution of two tagged density peaks. 
However, as argued in section \ref{can_correction}, the correction of the interaction at each time is equivalent 
to an equilibrium situation. So we can assess the relative magnitudes of the three terms in (\ref{can_correct_mix}) 
by considering a static situation for which the two tagged profiles are fixed to be overlapping 
Gaussians (see Fig.\ref{mix_corrections}a), although the detailed functional form chosen is not important. 

Choosing $A=I_1(\rb,t)$ from (\ref{av_force}) as target quantity in Eq.(\ref{can_correct_mix}), where particle $1$ is 
defined to be on the right, we numerically evaluate each of the three terms in the canonical correction 
series for a given separation between the Gaussian peaks (see Fig.\ref{mix_corrections}c). 
The three terms can be interpreted as the corrections to the force acting on particle one. $f_{1}^{11}$ corrects 
for grand-canonical fluctuations in $\rho_1$, $f_{1}^{00}$ for the increased interaction of 
$\rho_0$ with $\rho_1$ due to fluctuations in $\rho_0$, and $f_{1}^{10}$ for the interaction between fluctuations in $\rho_0$ and $\rho_1$.
If interactions are shortrange, it is clear that the term $f_{1}^{11}$ makes the dominant contibution to the first order correction 
and that the small term $f_{1}^{01}$ may be neglected to a good level of approximation. 

Just keeping the terms $f_{1}^{00}$ and $f_{1}^{11}$ corresponds to a first order canonical correction for each component separately. As we have shown in section \ref{singlerod} this corresponds to a first order suppression of the unphysical self-interaction, as each component consists of only one particle. Neglecting the mixed terms of the full canonical correction series leads to a system where each component does not interact with itself, but only with the other components.

The conclusion which we draw from this simple test is that a suppression of the self-interaction in a system of $N$ tagged 
density peaks, in which each peak represents a separate species and starts from sharp initial conditions, 
results in a reasonable approximation to the true Brownian dynamics at short and intermediate times.
However, the prohibitive numerical demands of performing the canonical transformation for more than a 
few particles (expecially if higher orders of correction are required) make desireable an alternative scheme 
for eliminating the self interaction.

\subsection{Eliminating the self interaction}

\subsubsection{Widom-Rowlinson model}
\label{sec:numerical}
A well established model for which particles of the same species do not interact is the Widom-Rowlinson (WR) model 
\cite{widomrowlinson,bradervink}. An approximate density functional for the $m$-component WR model has been 
derived \cite{schmidt2001} which provides a reasonable description of both the bulk phase behaviour and structural 
correlations. 
We propose to identify each species of an $N$-component WR model with an individual particle. 
Employing the Widom-Rowlinson functional in this fashion guarantees the absence of self interactions, 
but treats the interactions in an approximate fashion as the functional is not exact. 
In equilibrium with finite particle number (e.g. in confined geometries) the Percus and tagged WR 
functionals will yield different results. 
Due to the absence of unphysical self interactions the equilibrium density profiles of the tagged WR 
model should lie closer to the results of Brownian dynamics simulation than those obtained from 
the Percus functional.

\begin{figure}
\hspace*{-0.5cm}
\includegraphics[width=8cm]{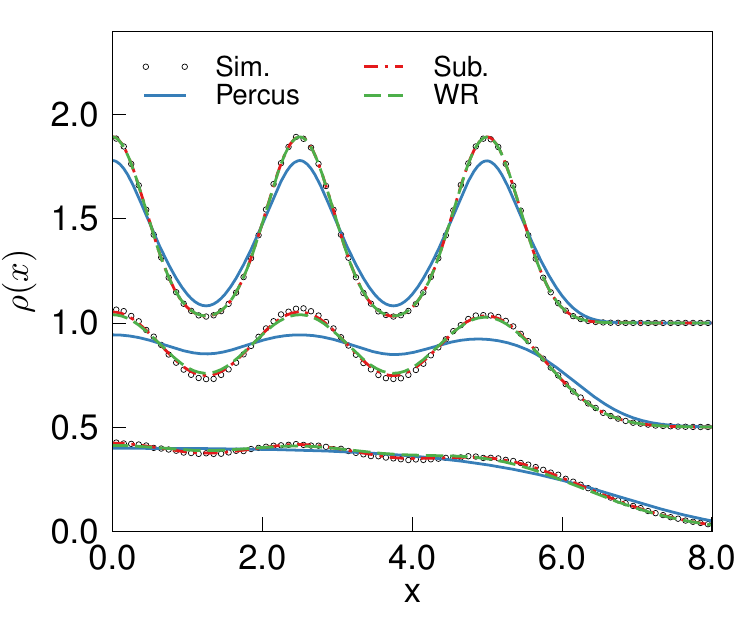}
\caption{\label{fig:expansion5} (Color online) 
Free expansion of five hard rods from a dense state at times $t=0.1,0.3,1.0$. 
For clarity, curves for $t=0.1$ and $0.3$ have been shifted upwards by $1$ and $0.5$, respectively, and we show only 
positive $x$ values.
The rods are initially located at $x=0,\pm2.5,\pm5$, corresponding to a volume fraction around $0.4$. 
Numerical results obtained from the Percus (\ref{percus_functional}), Widom-Rowlinson (\ref{wrfunctional}) and 
subtraction (\ref{subtraction}) functionals are compared with the results of Brownian dynamics simulation. The 
simulation errorbars in this and all subsequent figures are smaller than the symbols themselves. The unit length is set by the rod length $d$, the times are in units of $d^2/D_0$.
}
\end{figure}

In one dimension the approximate excess free energy of the WR model is given by \cite{schmidt2001}
\begin{align}\label{wrfunctional}
\Fex^{wr}(\set{\rho_i}) &= \int \mathrm{d}x \Phi(x)	&	\Phi &= \sum_{i=0}^m n_0^i \phi_i
\end{align}
with weighted densities $n_0$ and $n_3$ for each component of the mixture. $\phi_i$ are the first derivatives of the 0D excess free energy w.r.t. $n_3^i$ and given by
\begin{equation}
\phi_i = \ln\bigg(1 - m + \sum_{j=1}^m \exp(z_j)\bigg) - z_i.
\end{equation}
The fugacities $z_i$ can be calculated from the implicit equation
\begin{equation}
z_i \exp(z_i) = \bigg(1 - m + \sum_{j=1}^m \exp(z_i)\bigg) n_3^i
\end{equation}
The one particle direct correlation function required to calculate the average interaction force 
(\ref{av_force}) is obtained by functional differentiation of the excess free energy.

%, which is required for all DFT and DDFT calculations can be calculated to
%\begin{equation}
%\dcf_i(x) = - \int \d{x'} \phi_i(x') \omega_0^i(x - x') + \sum_{j-1}^m n_0^j(z') \phi_{ji}(x') \omega_3^i(x-x')
%\end{equation}
%The $\phi_{ik}$ are the second derivatives of the 0D excess free energy w.r.t. $n_3$ and given by
%\begin{equation}
%\phi_{ik} = \frac{\sum_{j=1}^m \exp(z_j) \partial_{n_3^k} z_j}{1 - m + \sum_{j=1}^m \exp(z_j)} - \partial_{n_3^k} z_i
%\end{equation}
%The derivatives of the fugacities can be calculated from another implicit equation
%\begin{equation}
%\begin{split}
%\exp(z_i)(\partial_{n_3^k} z_i)(1+ z_i) =& \delta_{ik}\bigg(1 - m + \sum_{j=1}^m \exp(z_j)\bigg) \\
%&+ n_3^i\bigg(\sum_{j=1}^m exp(z_j) \partial_{n_3^k} z_j\bigg)
%\end{split}
%\end{equation}

\subsubsection{Subtraction of the self energy}
A simpler, albeit {\em ad hoc} method of suppressing the self interaction 
is to first calculate the excess free energy of full $N$ component mixture using the Percus 
functional (\ref{percus_functional}) and then subtract the individual excess free energy of each density peak. 
The remainder thus provides an approximation to the desired excess free energy arising 
from interaction between different species. 
This prescription amounts to employing the `subtraction' functional 
\begin{equation}\label{subtraction}
\Fex^{sub}(\set{\rho_i}) = \Fex(\set{\rho_i}) - \sum_i \Fex(\rho_i)
\end{equation}
in the DDFT equation (\ref{stDDFTmix}).
While the ansatz (\ref{subtraction}) is not justified from fundamental principles, it nevertheless has a 
certain physical appeal. In particular $\Fex^{sub}$ vanishes for the case 
of a single particle and becomes exact for many particles in the low density limit. 
A similar approach was taken in \cite{hopkins2010}, in which an explicit self interaction term of a 
tagged particle was omitted from the Ramakrishnan-Yussouff functional 
\cite{ramakrishnan} in order to recover the exact single particle diffusion. 

\begin{figure}
\hspace*{-0.5cm}
\includegraphics[width=8cm]{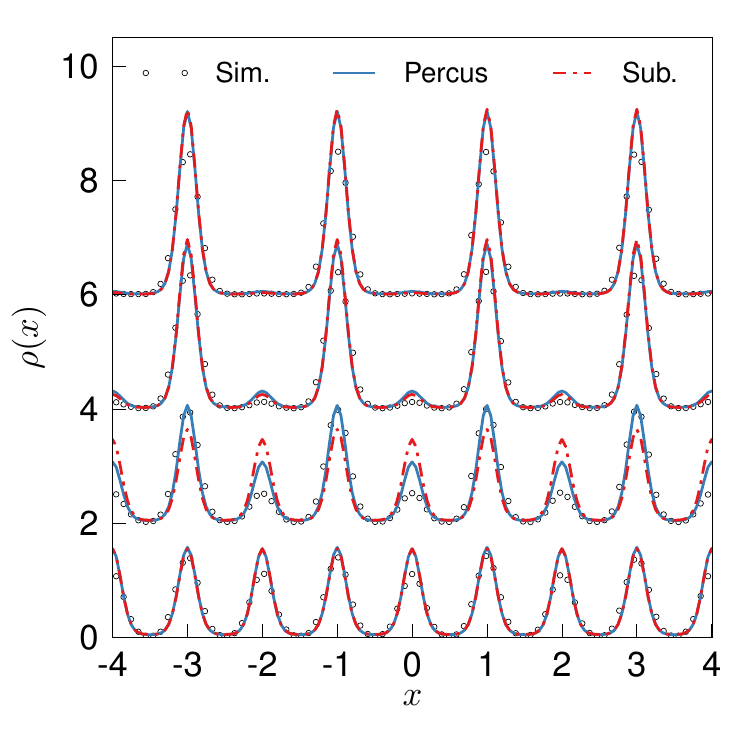}
\caption{\label{fig:highly-correlated} (Color online)
Evolution of the total density for the relaxation of four rods of length $d=1.6$ in a periodic potential with minima at integer values and periodic boundary conditions. The lengthscale is given by the period $h$ of the external potential, the timescale by $h^2/D_0$.
Shown are the results from simulations and DDFT calculations using the Percus functional (\ref{percus_functional}) 
and the subtraction functional (\ref{subtraction}) for times $t=1.0,5.0,25.0,100.0$. 
The corresponding evolution of a tagged density field is shown in Fig.\ref{fig:highly-correlated2}.
}
\end{figure}

\section{Numerical results for several rods}\label{numerical_results}

\subsection{Free Expansion}
\label{sec:freeex}
In order to compare the dynamics generated by the WR functional (\ref{wrfunctional}) with those generated by the subtraction functional (\ref{subtraction}) we consider the free diffusion of five rods from sharp initial 
conditions. 
The initial delta function distributions are each separated by $2.5$ particle diameters (corresponding to a 
volume fraction of around $0.4$). 
This choice ensures that the density fields have not entered the low density regime for the intermediate times 
at which overlap between neighbouring density peaks becomes significant. 
The present test, which is similar to that considered in \cite{marconitarazona99}, thus constitutes a genuine test 
of the functionals (\ref{wrfunctional}) and (\ref{subtraction}) beyond the second virial level. 
In Fig.\ref{fig:expansion5} we show the time evolution of the total density 
\begin{eqnarray}
\rho(x,t)=\sum_{i=1}^m\rho_i(x,t),  
\end{eqnarray}
with $m=5$, for three different times generated using the Percus (\ref{percus_functional}), WR (\ref{wrfunctional}) 
and subtraction (\ref{subtraction}) functionals in the multi-component DDFT equation (\ref{stDDFTmix}). 
 
As was the case for an isolated particle (see Fig.\ref{fig:one-particle-diff}), the short time relaxation of the 
density peaks generated by the Percus functional is too fast when compared with the Brownian dynamics simulation data.  
In contrast, the relaxation of the total density predicted by both the WR and subtraction functionals is in almost perfect agreement with the simulation results. 
This good level of agreement supports our argument that the self interaction is the primary source of error induced 
by a grand-canonical generating functional, at least for short and intermediate times.  
Close inspection of the data reveals that the subtraction functional describes the simulation data slightly better 
than the WR functional, but the difference is marginal. However, if the initial packing of the rods is denser, so that larger local densities occur, the WR functional becomes less reliable and predicts a unphysically fast relaxation, while the subtraction functional still gives a good description of the simulation data.

\subsection{Relaxation through a highly correlated state}
\label{sec:correlated}
We now consider a further test case for which rods with a length of $1.6$ relax from sharp initial conditions 
in a periodic external potential $\beta V_{\rm ext}(x)=-2\sin(2\pi x)$. The initial positions are chosen such that a particle is located at every second 
potential minimum.  
This situation was originally suggested in \cite{marconitarazona99} as a toy model for the study of slow 
dynamics.
The external potential and rod length are constructed in such a way that two rods cannot be 
simultaneously at the bottom of neighbouring potential minima.  
Transport of particles between neighbouring minima thus requires a correlated motion of all $N$ 
rods and it is the rarity of such events which leads to a long relaxation time. 
For long times an equilibrium solution is reached in which each potential well is equally populated. 

In Fig.\ref{fig:highly-correlated} we compare the time evolution of the total density from standard DDFT 
employing the Percus functional (\ref{percus_functional}) and from a tagged particle calculation 
employing the subtraction functional (\ref{subtraction}) with Brownian dynamics data.

\begin{figure}
\hspace*{-0.5cm}
\includegraphics[width=8cm]{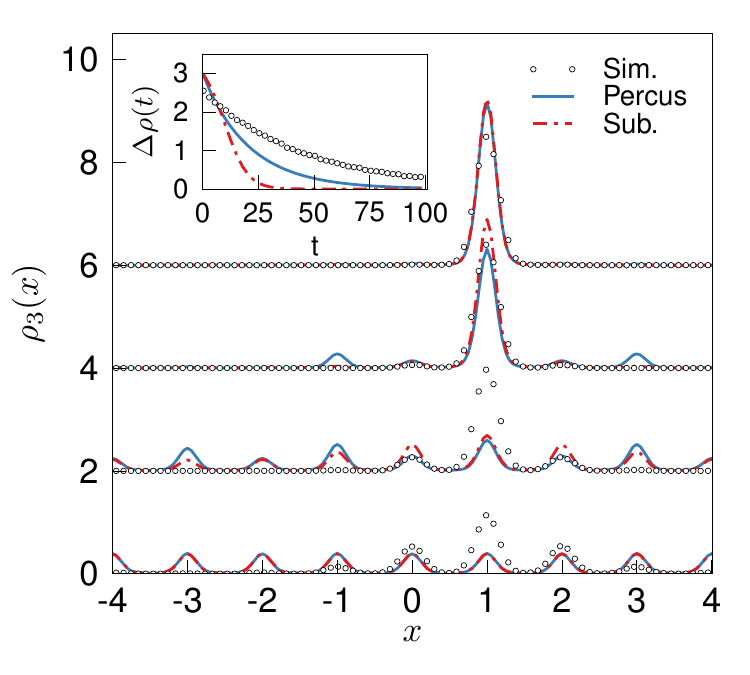}
\caption{\label{fig:highly-correlated2} (Color online)
A tagged density for same system considered in Fig.\ref{fig:highly-correlated} at $t=1.0,5.0,25.0,100.0$. 
The Percus functional generates an unphysical growth of density peaks in next-nearest-neighbour minima which 
is suppressed when employing the subtraction functional. 
The inset shows the difference in the total density between neighbouring potential minima. Units as in Fig. \ref{fig:highly-correlated}}
\end{figure}

For short times the relaxation generated by the Percus functional is faster than that of the 
subtraction functional, in keeping with the intuition obtained from the case of a single free particle 
(see section \ref{singlerod}). Surprisingly, for later times the relaxation of the subtraction functional profile 
overtakes that of the Percus profile, arriving more quickly to the equilibrium distribution in which, on average, 
half a rod is located in each well. 
Use of the Percus functional thus yields better agreement with the simulation data than the, supposedly superior, 
subtraction functional. 
However, the good performance of the standard theory arises from a fortuitous cancellation of errors, which can be appreciated by 
looking at the time evolution of a single tagged density peak. 

In Fig.\ref{fig:highly-correlated2} we show the time evolution of the tagged density corresonding to the third rod 
(labelling from left to right in Fig.\ref{fig:highly-correlated}). 
The other tagged densities evolve identically with time. 
Inspection of the profiles for $t=5$ reveals the strange behaviour of the tagged density from the Percus functional. 
Physically, it is reasonable to expect that the density peak will first diffuse into the neighbouring 
wells before spreading further to the next-nearest-neighbours, and so on. 
However, the Percus DDFT predicts that the density in the next-nearest-neighbour well grows more rapidly than that 
in the neighbouring well. 
For intermediate times, the unphysically large density which has built up in the two next-nearest-neighbour wells then 
pushes back on the central peak and slows its decay (i.e. generates a component of the self interaction which tends 
to confine the remaining density in the central peak). 

In contrast to the Percus functional, the subtraction functional has a strongly reduced self interaction and does not 
suffer from an unphysical decay of the tagged density. 
Unfortunately, the improved description provided by use of the subtraction functional destroys the error cancelation 
presented by the Percus functional profile and thus relaxes much faster than the simulation data.  
The relative relaxation rate of the two approaches can be appreciated from the inset to Fig.\ref{fig:highly-correlated2} 
which shows the difference in total density between neighbouring potential minima. 

The important message which emerges from this test case is that a qualitatively good description of 
the total density relaxation is not sufficient to conclude that the DDFT is really capturing correctly 
the underlying dynamics of the tagged density profiles. 

\begin{figure}
\hspace*{-0.8cm}
\includegraphics[width=8.4cm]{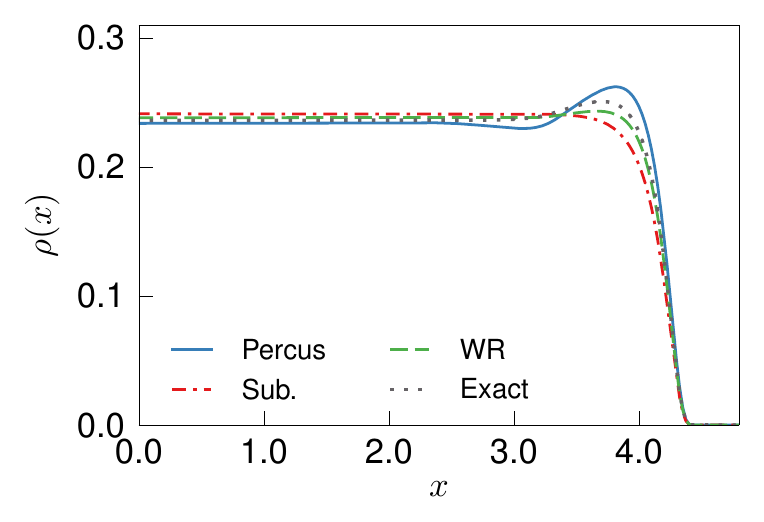}
\caption{\label{fig:static} (Color online)
Equilibrium total density profiles of two hard rods confined to a slit with soft walls (Yukawa potential with screening length 0.2) located at positions $x=\pm5$. The results are obtained from DFT Calculations with the Percus functional, the subtraction functional and the two component Widom-Rowlinson functional. The unit length is set by the rod length.}
\end{figure}

\subsection{Cage dynamics}
In two recent papers Hopkins {\em et al.} \cite{hopkins2007,hopkins2010} have shown that standard grand-canonical DDFT (in three-dimensions) can reproduce the two-step relaxation of the self part of the van Hove function 
\cite{hansenmcdonald86} characteristic of glass forming systems. 
At sufficiently high volume fractions a divergent relaxation time was identified. 
In these studies the one-component Gaussian-core \cite{hopkins2007} and hard-sphere \cite{hopkins2010} models were 
formally viewed as a two component mixture consisting of $N-1$ particles of species $d$ and a single particle of 
species $s$, thus tagging an arbitrary particle. 
A sharp initial condition was taken for particle $s$ and the corresponding 
equilibrium distribution for the density field of the $d$-component
\begin{eqnarray}
\rho_s(r,0)&=&\delta(r)\\
\rho_d(r,0)&=&\rho_b g(r),
\end{eqnarray}
where $\rho_b$ is the bulk density. 

Interpretation of the results of \cite{hopkins2007,hopkins2010} was complicated by the fact that an 
approximate quadratic density functional was employed in the DDFT equation. 
It remains an open question whether the observed slow dynamics, which arise from a minimum in the 
equilibrium free energy, are genuinely associated with the glass transition, or rather an indication 
of freezing within the theory.  
We think that some light can be shed on this issue by considering the very simple case of two rods confined between 
impenetrable walls. 
The unique ordering of the rods on the line restricts the kinetically accessible phase space. 
As this is a characteristic feature of arrested states in general, we believe that a robust account of 
the equilibrium tagged density distributions of two rods confined to 
a slit is a neccessary pre-requisite for any DDFT claiming a realistic description of dynamic arrest in 
higher-dimensional systems.  

\begin{figure}
\hspace*{-0.8cm}
\includegraphics[width=8.4cm]{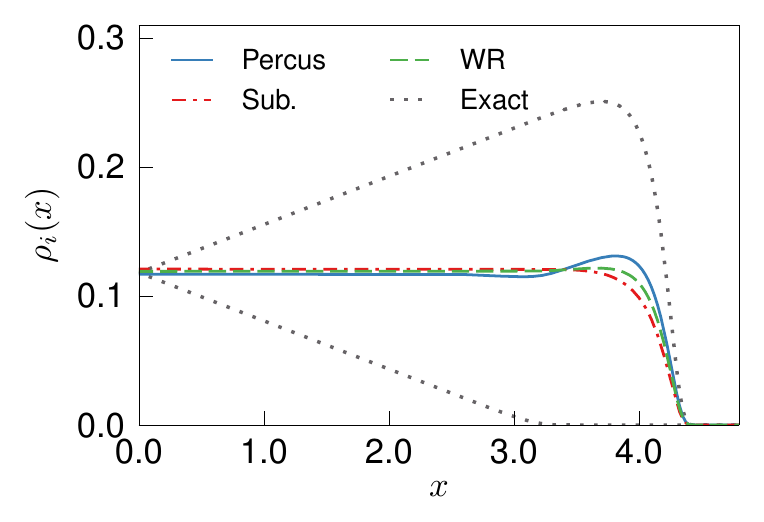}
\caption{\label{fig:no-cage} (Color online)
The tagged particle density for the situation of figure \ref{fig:static}. The exact density profile clearly shows how the particles are constrained to one side of the slit, as the particles cannot move through each other. 
At the wall, the exclusion zone for the left particle is clearly visible. 
The DFT profiles cannot capture this feature and the tagged particle densities of both particles are identical and 
delocalized over the whole slit. Units as in figure \ref{fig:static}}
\end{figure}

In Fig.\ref{fig:static} we compare the equilibrium total density from DDFT with simulation data for two rods 
confined by the potential
\begin{eqnarray}
\beta V_{\rm ext}(x)= \frac{e^{\tfrac{x - 4.5}{l}}}{x-4.5} + \frac{e^{\tfrac{4.5-x}{l}}}{4.5-x}
\end{eqnarray}
with decay length $l=0.2$.  
This gives a volume fraction around $0.2$. For this situation it can easily be shown that the canonical equilibrium density of the left particle is given by
\begin{equation}
\rho_1(x) = e^{-\beta V_{\rm ext}(x)} \int_{x + d}^\infty dx' e^{-\beta V_{\rm ext}(x')} 
\end{equation}

The primary observation to be made from Fig.\ref{fig:static} is that the Percus functional generates  
a packing structure at the wall with a well developed peak and a dip.
The dip is absent in the exact profile and the peak is less pronounced. 
It should be recalled that the Percus functional profiles represent the exact grand-canonical result for 
a slit with $\langle N\rangle=2$. The  peak is slightly underestimated by the WR functional 
(\ref{wrfunctional}) and entirely absent from the subtraction functional profile.
Overall the WR functional shows the best level of agreement with the simulation data 
for the total density $\rho(x)$.

On the basis of the total density profiles shown in Fig.\ref{fig:static} it is tempting to conclude 
that the WR functional provides an acceptable description of the equilibrium density distribution.
Indeed, this is true if one is interested only in the total density $\rho(x)$. 
A completely different picture emerges, however, when considering the two tagged density profiles. 
In Fig.\ref{fig:no-cage} we show the tagged densities corresponding to the same situation as shown in 
Fig.\ref{fig:static}. 
The difference between the theoretical predictions and the simulation data is dramatic. 
All of the DFT approaches yield identical equilibrium tagged profiles $\rho_1(x)=\rho_2(x)=\rho(x)/2$, clearly 
demonstrating that the DDFT time evolution does not respect the fact that hard-core particles cannot pass 
through each other.  
At some point during the time evolution the density fields always tunnel through each other to arrive at unphysical 
regions of phase space. 
Given that none of the functionals considered in the present work are capable of capturing this most elementary 
`caging' dynamics, we find it unlikely that the DDFT employed in \cite{hopkins2007,hopkins2010} is capable of 
capturing a true nonergodic transition. The constrained order of the rods is encoded in the hierarchy of correlation functions (see Section \ref{sec:fundamentals} \ref{ssec:ddft}) in a complicated way, so one can not expect that truncating this hierarchy preserves this property.

\section{Discussion}\label{discussion}
We have analysed the shortcomings of DDFT for the description of localised density distributions using 
a simple one-dimensional model of hard-rods. 
This model has the advantage that the exact grand-canonical equilibrium functional is known,  thus removing the 
possibility of dynamical artifacts arising from use of an approximate functional. 
It is well known among DDFT practitioners that the standard theory predicts a relaxational 
dynamics which is systematically too fast when benchmarked against Brownian dynamics simulation. 
By employing a formal expansion of the canonical average we conclude that this deficiency is due to the unphysical 
self interaction arising from grand-canonical coupling to a reservoir. 

In situations where few particles are strongly confined, the presence of a self interaction leads to equilibrium 
density profiles which differ from thise obtained in Brownian dyanmics simulation. 
We anticipate that the self interaction will become relevant for large systems in cases for which large unbalanced forces 
are present, e.g. during the relaxation of large density gradients in a nonequilibrium profile. 
The corrections in the force resulting from removal of the self interaction will modify the relaxation timescale and 
slow the relaxation relative to standard grand-canonical DDFT. 

The only method by which the self interaction can be removed is to individually tag the density field of each 
particle and then employ either the canonical expansion series, a 
Widom-Rowlinson-type model or {\em ad hoc} subtraction of undesirable contributions to the excess free energy 
functional. 
Of these three possibilities, the latter proved to be both the most reliable and simplest to implement.  
 
Once the self interaction has been dealt with appropriately, we find that the free expansion of any number 
of interacting rods can be well described using DDFT. However, the considered test case which focused on the collective 
motion of rods (see section \ref{sec:correlated}) demonstrated that the previously reported good agreement between 
DDFT and simulation \cite{marconitarazona99} is due to cancellation of errors. 
Removal of the self interaction corrects the unphysical aspects of the tagged density relaxation, with the 
consequence that the results for the relaxation of the total density become worse (even faster than standard DDFT). 
The fact that DDFT clearly has difficulty in describing the relaxation through highly correlated states raises 
suspicions about its ability to capture the slow dynamics of dense systems.  

Although our one-dimensional model provides only a crude description of a real fluid, it 
nevertheless enables some of the subtleties 
associated with a partitioned phase space to be investigated within a simple setting.  
Motivated by the recent work of Hopkins {\em et al.} \cite{hopkins2007,hopkins2010} we presented the case 
of two confined rods as a toy model for the cage effect, whereby particles in a glass are localised by the 
geometrical contraints of their neighbours. 
Given that the very simple geometrical contraints on the tagged particle densities are not respected 
(see Fig.\ref{fig:no-cage}) it would be remarkable if the same theory 
turned out to be capable of describing the spontaneous localization associated with the glass transition in 
two- and three-dimensions. 
One possibility is that the particular functional employed in \cite{hopkins2007,hopkins2010}, namely the 
quadratic Ramakrishnan-Yussouff functional \cite{ramakrishnan}, is somehow able to compensate for the 
errors arising from a grand-canonical treatment of the density. 
However, such a scenario would seem to require a highly nontrivial cancellation of errors.  

By considering test cases in which finite numbers of rods are confined to a slit we now strongly suspect that any 
theoretical approach which is closed on the level of the one-body density, such as DDFT, will be unable to 
describe localization of the tagged density fields (at least when employing the exact equilibrium functional). 
%(a view supported by the recent  Bethe ansatz solution of 
%this problem \cite{lizana}). 
When working solely with the one-body density, effective interactions between 
the average quantities $\rho_i({\bf r})$ are implemented, whereas Brownian dynamics considers interactions 
between the density operators $\hat\rho_i({\bf r})=\sum_i\delta({\bf r}-{\bf r}_i)$ before taking the 
average. 
This mean-field treatment of the interaction between tagged density fields appears to make the 
tunnelling of tagged density into geometrically forbidden regions unavoidable. 
We note that these limitations do not neccessarily pose a problem for the established density functional 
theory of freezing \cite{oxtoby91}. 
In such calculations the crystal is identified as a periodic variation of the total density field and no 
claim is made to identify any specific particle: the tagged densities are fully delocalized throughout the sample.

To arrive at a theory which respects the geometrical constraints on the tagged density fields it 
is neccessary to go beyond a density based description and consider explicitly the dynamics of the 
higher-order density correlators, i.e. going beyond the simplest adiabatic approximation. 
Indeed, the importance of improving upon the standard adiabatic approach was recently identified 
by Haataja {\em et al.} \cite{loewen_adiabatic} as one of the most important open problems in DFT. 
A fundamental question is whether a finite order truncation of the dynamical hierarchy 
(of which (\ref{eq:eom_rho1}) and (\ref{flux}) constitute the first member) can capture tagged density 
localization.  

Integration of the Smoluchowski equation over all but two of the particle coordinates leads to an equation 
of motion for the pair density involving a weighted integral over the three particle correlations. 
Specifically, one is required to evaluate integrals of the form
\begin{eqnarray}\label{integral}
\int \!d\rb_3\, \left[\frac{\rho^{(3)}(\rb_1,\rb_2,\rb_3)}{\rho^{(2)}(\rb_1,\rb_2)}\right](-\nabla_1\phi(r_{13})).
\end{eqnarray}
The factor in square brackets can be identified as the conditional probability to find a particle at $\rb_3$, 
given that there are particles at positions $\rb_1$ and $\rb_2$. 
The integral thus represents the average force acting on a particle at $\rb_1$ due to interactions mediated by 
the other particles (an analogous integral provides the indirect force on a particle at $\rb_2$).
Making an adiabatic approximation, the integral (\ref{integral}) may be replaced by 
$\nabla_1 c_{\,\rb_2}^{(1)}(\rb_1)$, 
where $c_{\,\rb_2}^{(1)}(\rb_1)$ is the one-body direct correlation function at $\rb_1$ calculated in the 
presence of both the physical external field and a test particle fixed at $\rb_2$ (hence the parametric dependence 
upon $\rb_2$). This leads to 

\begin{eqnarray}\label{eddft}
\frac{\partial\rho^{(2)}(\rb_1,\rb_2)}{\Gamma\partial t}
&=&\nabla_1\cdot\left[\, \rho^{(2)}(\rb_1,\rb_2)\nabla_1
\left(\frac{\delta\mathcal{F}}{\delta\rho(\rb_1)} 
\right)_{\!2}
\,\right] \notag\\
&+&
\nabla_2\cdot\left[\, \rho^{(2)}(\rb_1,\rb_2)\nabla_2
\left(\frac{\delta\mathcal{F}}{\delta\rho(\rb_2)} 
\right)_{\!1}
\,\right].\;\;\;
\end{eqnarray}
Combining Eqs.(\ref{eq:eom_rho1}), (\ref{flux}) and (\ref{eddft}) leads to a closed theory for the dynamics 
of the one- and two-body density in the form of a pair of coupled first order (in time) differential 
equations. 
A similar equation was employed in \cite{penna_tarazona2} (see their Eq.(9)) in which the three-body density 
was treated using a superposition approximation. 
Crucially, in (\ref{eddft}) the free energy functional to be differentiated contains an external 
field consisting of the physical external potential of interest, $V_{\rm ext}(\rb)$, {\em plus} 
a `test' particle held fixed at either position $\rb_1$ or $\rb_2$, as indicated by the subscript 
on the functional derivative. 
Equation (\ref{eddft}) is still `adiabatic', in the sense that 
three- and higher-body correlations are determined from the nonequilibrium 
$\rho(\rb,t)$ and $\rho^{(2)}(\rb_1,\rb_2,t)$ using equilibrium statistical mechanical relations. 
Nevertheless, this extended theory goes beyond (\ref{stDDFT}), as the pair density is no longer 
tied to the instantaneous value of the density and relaxes instead on a finite timescale.  
Note that the formal elimination of $\rho^{(2)}(\rb_1,\rb_2)$ from the coupled equations generates, in 
principle, an equation for $\rho(\rb_1)$ alone which includes memory effects \cite{footnote_zwanzig}. 
In practice, however, the nonlinearity of the equations does not allow an explicit form for the 
memory kernel to be obtained and memory effects remain implicit to the coupled system of equations. 

In equilibrium, Eq.(\ref{eddft}) predicts that the pair correlations are those obtained from a 
test-particle calculation using the chosen free energy functional. 
The exact single particle dynamics are recovered using the initial condition 
$\rho^{(2)}(\rb_1,\rb_2,0)\!=\!0$. 
More generally, the correct normalisation of the initial pair density, 
$\int d\rb_1\!\int d\rb_2\,\rho^{(2)}(\rb_1,\rb_2,0)\!=\!N(N-1)$, is preserved throughout the time evolution, 
which is not the case when applying the standard adiabatic approximation, i.e. calculating equilibrium 
pair correlations in the presence of the effective potential (\ref{eq:Veff}). 
Although we defer a more detailed investigation of (\ref{eddft}) to future work, preliminary calculations 
indicate the results for the nonequilibrium one- and two-body density of hard-rods are considerably 
improved, relative to standard DDFT. Of particular interest will be the predictions of (\ref{eddft}) for 
the density profile of confined systems in the limit of long times.  
  
Apart from our preliminary investigations of (\ref{eddft}) the only explicit test of the adiabatic 
approximation of which we are aware was performed using 
continuous-time 
Monte-Carlo simulations of the Potts model subject to Glauber dynamics \cite{heinrichs03}. 
In this study, initial simulation configurations were prepared which reproduced the equilibrium one-body 
occupation number profile, but with nonequilibrium correlations between the occupation numbers. 
During the simulated time-evolution the relaxation of higher-order correlation functions caused the the one-body 
profile 
to drift first out of equilibrium, passing through a sequence of transient intermediate states, before returning 
back to equilibrium at long times. 
These findings clearly cannot be reproduced by theories based on $\rho(\rb_1,t)$ alone,  
as no distinction can be made  between states with the same one-body profile but different higher-body correlations. 
This issue may prove to be important when considering systems with slow dynamics for which the one-body 
density remains constant during gradual structural relaxation processes. 

We thank M. Schmidt and A.J. Archer for stimulating discussions. Funding was provided by the Swiss National 
Science Foundation.

\bibliography{all}{}

\end{document}